\documentclass{elsart}
\usepackage{epsfig}
\usepackage{amsmath}
\usepackage{bbm}
\usepackage{pifont}
\usepackage{pstricks,pst-node}
\newcommand{\ket}[1]{\ensuremath{|#1\rangle}}

\newcommand{\braket}[2]{\ensuremath{\langle #1|#2\rangle}}

\newcommand{\KetBraD}[2]{\ensuremath{| #1 \rangle\langle #2 |}}

\newcommand{\hil}{\ensuremath{\mathcal{H}}}

\newcommand{\BE}{\begin{equation}}
\newcommand{\EE}{\end{equation}}
\newcommand{\be}{\begin{equation}}
\newcommand{\ee}{\end{equation}}
\newcommand{\bea}{\begin{eqnarray}}
\newcommand{\eea}{\end{eqnarray}}
\newcommand{\kommentar}[1]{}

\newcommand{\ew}[1]{\ensuremath{\langle #1 \rangle}}

\newcommand{\eg}{{\it e.g.}}
\newcommand{\via}{{\it via\ }}

\newcommand{\beq}{\begin{equation}}
\newcommand{\eeq}{\end{equation}}
\newcommand{\beqa}{\begin{eqnarray}}
\newcommand{\eeqa}{\end{eqnarray}}
\newcommand{\bse}{\begin{subequations}}
\newcommand{\ese}{\end{subequations}}

\newcommand{\eqnref}[1]{Eq.~\eqref{#1}}

\newcommand{\figref}[1]{Fig.~\ref{#1}}

\begin{document}

\title{Three level atom optics in dipole traps and waveguides}

\begin{frontmatter}
\author[bcn,hannover]{K. Eckert},
\author[bcn]{J. Mompart},
\author[bcn]{R. Corbal{\'a}n},
\author[icfo,hannover]{M. Lewenstein\thanksref{icrea}}, and
\author[darms]{G. Birkl}
\address[bcn]{Departament de F\'{\i}sica, Universitat Aut\`onoma de Barcelona,  E-08193 Bellaterra, Spain.}
\address[icfo]{ICFO - Institut de Ci\`encies Fot\`oniques, 08034 Barcelona, Spain.}
\address[darms]{Institut f\"ur Angewandte Physik, Technische Universit\"at Darmstadt, Schlossgartenstra{\ss}e 7, D-64289 Darmstadt, Germany.}
\address[hannover]{Institut f{\"u}r theoretische Physik, Universit{\"a}t Hannover, D-30167 Hannover, Germany.}

\thanks[icrea]{also at Instituci\'o Catalana de recerca i estudis avan\c cats.}

\begin{abstract}
An analogy is explored between a setup of three atomic traps coupled via tunneling and 
an internal atomic three-level system interacting with two laser fields. Within this
scenario we describe a STIRAP like process which allows to move an atom
between the ground states of two trapping potentials and analyze its robustness.
This analogy is extended to other robust and coherent transport schemes and to systems
of more than a single atom. Finally it is applied to manipulate external degrees
of freedom of atomic wave packets propagating in waveguides.
\end{abstract}
\end{frontmatter}

\maketitle

\section{Introduction}

Exploring the wave nature of massive particles has become possible through
the enormous experimental advances in the cooling of neutral atoms, ions, and
molecules to temperatures where the de Broglie wavelength becomes comparable
to or larger than optical wavelengths. These achievements have stimulated great interest
into the field of quantum {\it atom} optics as an analogue of quantum optics
with light \cite{b:meystre,rolston:2002}. A major objective within this field is to develop
elements for the manipulation of the spatial wavefunction of atoms or atomic ensembles, as
beam splitters, mirrors, lenses, etc. Applications are broad, ranging from a fundamental
interest in probing the wave nature of particles to the manipulation of neutral atoms
for implementing quantum gates and to the construction of atom interferometers for
precision measurements of physical constants or as inertial sensors. In all these cases,
a crucial requirement is  -- as in quantum optics -- to preserve the coherence of the matter wave. 

Of special interest to atom interferometers as well as to quantum information processing
are concepts to trap or to guide atomic
matter waves. For trapped atoms, the interaction with external lasers
can be precisely controlled, a spreading of the wave packet can be inhibited
in some or all spatial dimensions, and the effect of gravity can be compensated.
Trapping and guiding of neutral atoms is usually based either on the interaction of
the atom's permanent magnetic dipole moment with magnetic fields \cite{reichel:2001}
or on the coupling of laser fields to the atom's induced dipole moment \cite{grimm:2000}.
Arrangements of current-carrying wires \cite{folman:2002,reichel:2001}, superpositions of
standing light-waves with different frequencies (superlattices) \cite{guidoni:1999},
or appropriately shaped
microlenses \cite{birkl:2001,dumke:2002} allow to design and control a variety of potential shapes.
Examples are Y-shaped guiding geometries to split a wave packet \cite{stenholm:2003},
cold atoms storage rings from guides forming a closed loop \cite{reichel:2001,folman:2002}, or traps
whose separation can be controlled in time \cite{dumke:2002,bergamini:2004,schumm:2005}. As two traps are brought to a
close distance and tunneling takes place, an atom initially located in one of them
oscillates in a Rabi-type fashion between the two potentials. This is in close resemblance
to a two-level atom interacting with a laser field, but in contrast the 'Rabi-frequency' is controlled
via tuning the tunneling interaction. Such a process, if implemented correctly,
is coherent as it does not introduce uncontrollable phases, and it indeed allows for
a simple realization of quantum bits and quantum gates \cite{mompart:2003}. This technique however is not
robust under variations of the system parameters and thus requires precise temporal control 
of the potentials. The same problem is of course present in optical two-level system,
where for this reason a variety of robust techniques have been developed, which
are based on controlling couplings in multi-level systems, and which are nowadays 
standard techniques in experiments.

Here we will provide a theoretical analysis 
of atom optical analogues to three-level techniques, especially discussing
processes reminiscent of stimulated Raman adiabatic passage (STIRAP, \cite{bergmann:1998})
to coherently move atoms between traps and and coherent population trapping
(CPT, \cite{arimondo:1996}) to create spatial superpositions of atomic wavefunctions.
We will furthermore provide simulations showing that this technique is
not only applicable to trapped atoms, but also to wave packets propagating in guiding potentials.

\section{Time-dependent trapping potentials}

\begin{figure}[tbp]
\begin{center}
\includegraphics[width=0.9\columnwidth]{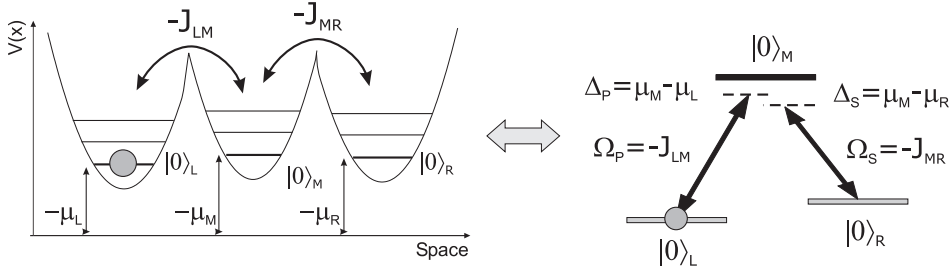}
\end{center}
\caption[]{
Illustration of the analogy between a system of three coupled trapping potentials arranged
linearly and an atomic three-level system in $\Lambda$-configuration. The tunneling matrix elements
$J$ correspond to the optical Rabi frequencies, the detunings are given by the difference
between the on-site energies.
}
\label{fig:tlsystem}
\end{figure}

To obtain an analogy between external degrees of freedom of an atom in a system of
trapping potentials coupled via tunneling and an electronic three-level system coupled
via the electric dipole-interaction with two laser fields, consider a linear arrangement
of three atom traps. We assume strong confinement in the orthogonal directions,
such that the dynamics can be restricted to the one-dimensional (1D) Hamiltonian
\be
\hat\hil_{\rm free}=\int dx\,\hat{\psi}^{\dagger}(x)\left[\frac{p_x^2}{2m}+V(x,t)\right]\hat{\psi}(x)
\equiv\int dx\,\hat{\psi}^{\dagger}(x)H_{\rm free}\hat{\psi}(x)
,
\label{eq:hamfree}
\ee
where $V(x,t)$ describes the potential consisting of three traps with tunable distance.
At each time $t$, $H_{\rm free}$ can be diagonalized to obtain the instantaneous eigenstates.
For large distance between the traps, these are localized at the
centers of the traps. In the general case, states $\phi_L(x,t)$, $\phi_M(x,t)$,
and $\phi_R(x,t)$, localized around the centers of the left, middle, and right trap, respectively,
can be constructed from suitable combinations of the eigenstates with lowest energy. 
Restricting to these states, we can expand $\hat{\psi}(x)=\sum_{\alpha=L,M,R}\hat{b}_{\alpha}\phi(x)$
to obtain \cite{jaksch:1998}
\be
\hil_{\rm free}=-J_{LM}(t)\hat{b}_L^{\dagger}b_M-J_{MR}(t)\hat{b}_M^{\dagger}b_R-
\frac12\sum_{\alpha=L,M,R}\mu_{\alpha}(t)\hat{b}_{\alpha}^{\dagger}\hat{b}_{\alpha}+{\rm c.c.}
\label{eq:ham_a}
\ee
Here $J_{\alpha\beta}(t)=-\int dx\phi_{\alpha}^*(x,t)H_{\rm free}(x,t)\phi_{\beta}(x,t)$
describes nearest-neighbor tunneling and $\mu_{\alpha}(t)=-\int dx\phi_{\alpha}^*(x,t)H_{\rm free}(x,t)\phi_{\alpha}(x,t)$
are the on-site energies. Interactions of next-nearest neighbors have been neglected.
Considering just a single atom and shifting the ground state energy, we arrive at the
following Hamiltonian isomorphic to the Hamiltonian of a three-level system coupled by two
laser fields in the rotating wave approximation \cite{rwa:????,bergmann:1998}:
\bea
H&=&-J_{LM}(t)(\KetBraD{L}{M}+\KetBraD{M}{L})-J_{MR}(t)(\KetBraD{M}{R}+\KetBraD{R}{M})\nonumber\\
&&-(\mu_M(t)-\mu_L(t))\KetBraD{M}{M}-(\mu_R(t)-\mu_L(t))\KetBraD{R}{R}.\label{eqn:ham_lambda}
\eea
The couplings $-J_{LM}$ and $-J_{MR}$ are the analogies of the Rabi frequencies of the pump and the Stokes 
laser, and $\mu_M-\mu_L\neq0$ and $\mu_R-\mu_L\neq0$ correspond to the detuning from the 
single- and two-photon transition, respectively (compare \figref{fig:tlsystem}).

Optical three-level systems have been extensively analyzed.
Exploiting the different possible configurations of detunings and variations
of the Rabi frequencies gives rise to a large number of coherent manipulation schemes
of the underlying three-level system, among them 
stimulated Raman adiabatic passage (STIRAP, \cite{bergmann:1998}), coherent population
trapping (CPT, \cite{arimondo:1996}), and
electromagnetically induced transparency (EIT, \cite{harris:1997,marangos:1998}). 
The analogy to the system of three coupled
traps, as demonstrated by the Hamiltonian \eqnref{eqn:ham_lambda}, clearly suggests
to explore the application of these effects to coherently manipulate the {\it external}
degrees of freedom of a trapped neutral atom, given the ability to control
the corresponding trap parameters. Such a control should be possible in various trapping
configurations as optical \cite{birkl:2001,dumke:2002,schlosser:2001} and magnetic
\cite{reichel:2001,folman:2002,schumm:2005} microtraps as well as in optical lattices
by exploiting superlattice techniques \cite{guidoni:1999}.

In the following we will especially refer to  
neutral atoms trapped in arrays of optical microtraps created by illuminating a set of
microlenses with a red detuned laser beam \cite{dumke:2002},
such that in each of the foci of the individual lenses neutral
atoms can be stored by the dipole force. By illuminating the microlenses by
independent laser beams under different angles, it is possible
to generate various sets of traps which can be approached or separated by changing the angle between the two lasers
\cite{dumke:2002}. Changing the angle allows to control the couplings between the traps, the on-site
energy can be tuned by changing the laser intensities. The optical potential
generated by a single laser passing through a single lens has a gaussian shape, i.e.,
\begin{equation}
V_{\rm trap}(x)=-V_0\exp\left({-\frac1{2V_0}m\omega_x^2x^2}\right)=-V_0\exp\left({-\frac{\hbar\omega_x}{2V_0}(\alpha x)^2}\right),
\label{eq:pot}
\end{equation}
where $V_0$ and $\omega_x$ are depth and frequency of the potential, respectively, $m$
is the mass of the neutral atom and $\alpha^{-1}=\sqrt{\hbar/m\omega_x}$ is the ground
state size. As several potentials of this type are superimposed, controlling the coherent
and adiabatic evolution of a trapped atom is complicate as the potential depth doubles
if two traps lie on top of each other. To avoid this problem, either techniques
to suppress non-adiabatic excitations can be employed \cite{haensel:2001,mompart:2003},
or the light intensity can be reduced in an appropriate way as the traps are approached, e.g.,
by adding a blue detuned laser to produce a compensating extra potential \cite{helge:2004}.
Here we will focus on the later option and assume $V(x)={\rm min}[V_{\rm trap}(x+a_L(t)),V_{\rm trap}(x),
V_{\rm trap}(x-a_R(t))]$, where $a_{L}(t)$ and $a_{R}(t)$ fix the centers of the traps. For
$m^2\omega_x^4a_{L/R}^2/(8V_0)\ll1$ the potential then consists of three concatenated harmonic
traps. 

The Hamiltonian (\ref{eqn:ham_lambda}) neglects contributions from non-adiabatic couplings to
excited vibrational states as well as direct couplings from the left to the right trap. In what
follows we will take into account the full Hamiltonian (\ref{eq:hamfree}) through
a numerical integration of the 1D Schr{\"o}dinger equation to simulate the
dynamics of a neutral atom in the three-trap potential.

\subsection{STIRAP -- robust shifting of atoms between traps}

\begin{figure}[tbp]
\begin{center}
\includegraphics[width=\columnwidth]{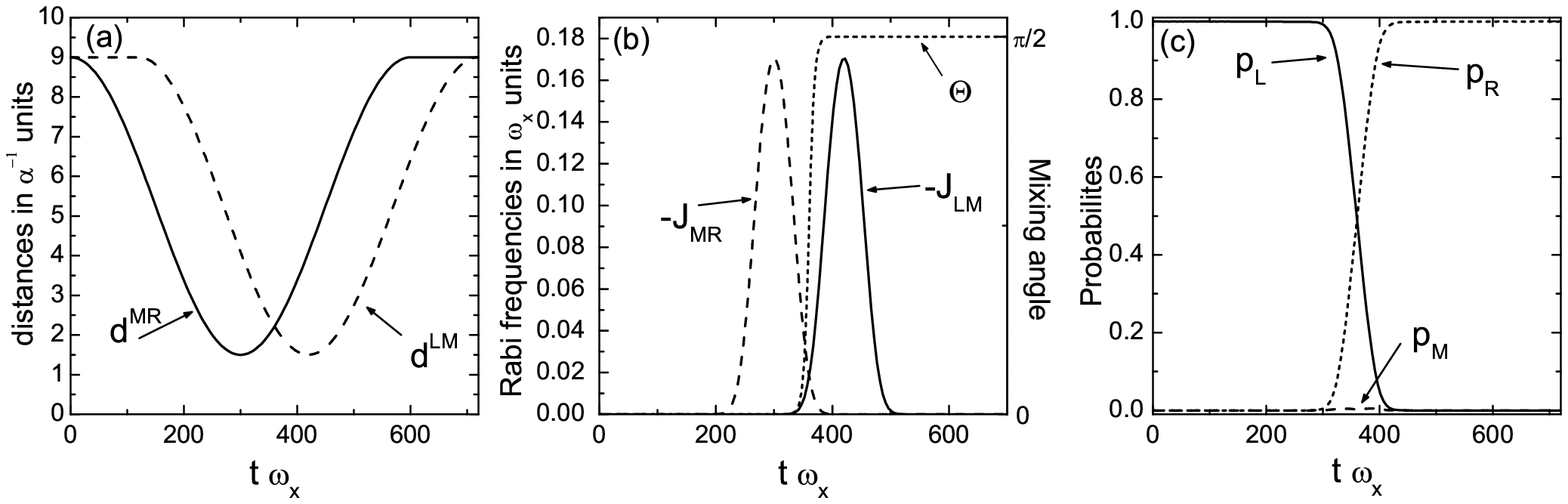}\\
\includegraphics[width=0.8\columnwidth]{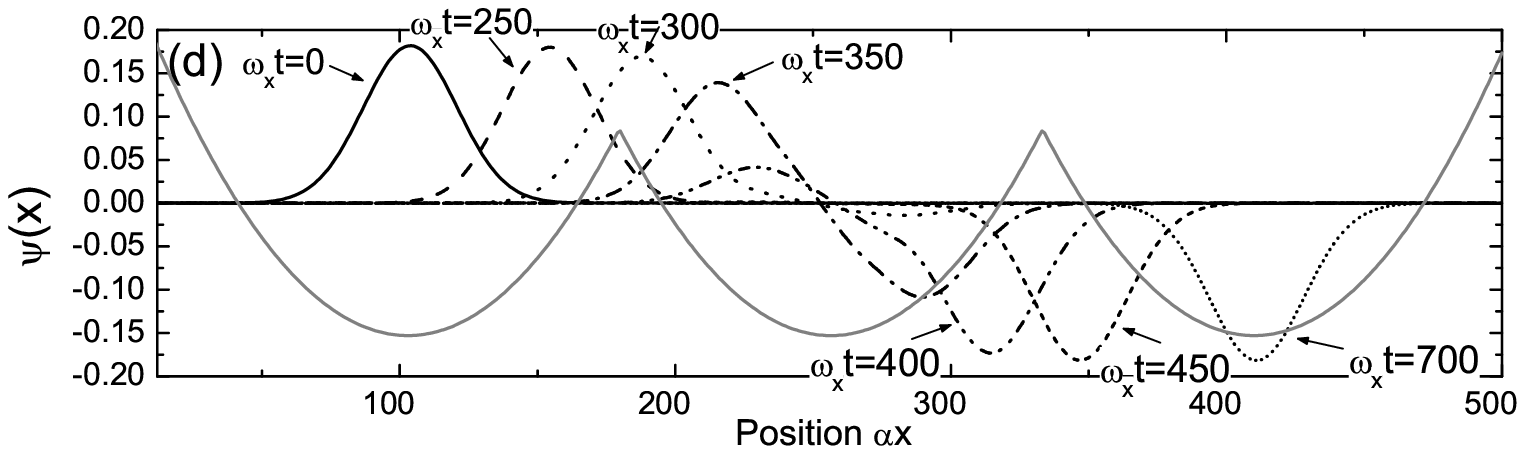}
\end{center}
\caption[]{
(a) Approaching sequence for a STIRAP-like process, 
(b) the evolution of the tunneling Rabi frequencies $-J_{LM}(t)$
and $-J_{MR}(t)$ between the left and the middle and the middle and
the right trap, calculated from the tunneling splitting energy of two traps, together with the mixing angle $\Theta(t)=\arctan\left(J_{LM}(t)/J_{MR}(t)\right)$,
and (c) the corresponding ground state populations; 
The parameters are
$d^{LM}_{max}\alpha=d^{MR}_{max}\alpha=9$ (maximal distances between traps), 
$d^{LM}_{min}\alpha=d^{MR}_{min}\alpha=1.5$ (minimal distances),
$t^{LM}_{r}\omega_x=t^{MR}_{r}\omega_x=300$ (time used to approach/separate the traps),
$t^{LM}_{i}\omega_x=t^{MR}_{i}\omega_x=0$ (time at which the traps are at the minimal distance,
and $t_{delay}\omega_x=120$ (delay between the the approaching processes).
(d) The spatial wavefunction $\braket{x}{D}$ of the dark state for various
times. The gray line gives the potential at $\omega_x t=0$.
}
\label{fig:tlao:stirap:single}
\end{figure}

For zero detunings, one of the eigenstates of Hamiltonian (\ref{eqn:ham_lambda}), the
{\it dark state}, only involves the states localized in the left and the right trap:
\beq
\ket{D(\Theta)}=\cos\Theta\ket{L}-\sin\Theta\ket{R}.
\eeq
Here $\Theta$ is the mixing angle which depends on the couplings through
$\cos\Theta=J_{LM}/J_{MR}$. To implement a robust method to move an
atom from the leftmost to the rightmost trap using tunneling, the counter-intuitive
STIRAP sequence can be applied: first the right and
middle traps are approached and separated, and, with an appropriate delay time $t_{\rm delay}$, the
same sequence is used for the left and middle trap [\figref{fig:tlao:stirap:single} (a)].
This changes the mixing angle from $\Theta=0$ to $\Theta=\pi/2$
[\figref{fig:tlao:stirap:single} (b)], and if the atom initially is located in the
left trap and the process is adiabatic, then the state is at all times
identical to the dark state. This moves the atom directly from $\ket{L}$
to $\ket{R}$ [\figref{fig:tlao:stirap:single} (c)]. The spatial wavefunctions $\braket{x}{D}$
of the dark state for various times during the approaching process are plotted in
\figref{fig:tlao:stirap:single} (d).

\begin{figure}[tbp]
\begin{center}
\includegraphics[width=\columnwidth]{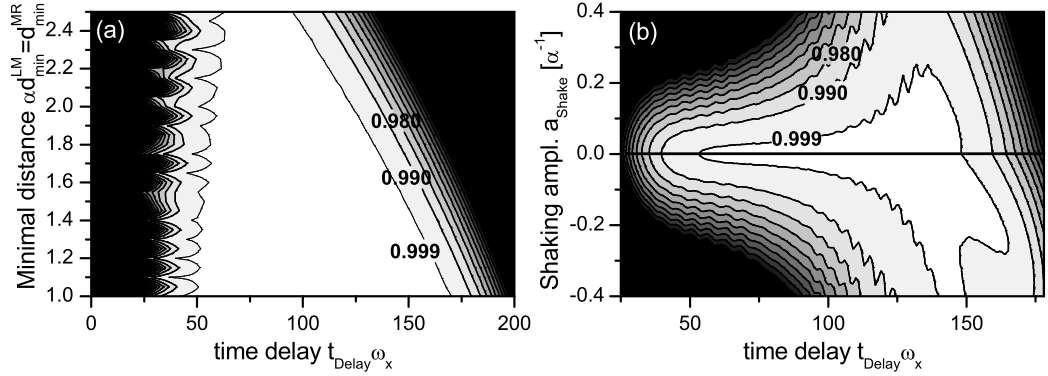}
\end{center}
\caption[]{
Robustness of the atom optics version of STIRAP, i.e., the transfer efficiency 
from $\ket{L}$ and $\ket{R}$, measured by the population $\rho_R=|\braket{R}{\psi(t)}|^2$.
All parameters not varied in the figures are
as in \figref{fig:tlao:stirap:single}.
In (a) the delay time $t_{\rm delay}$
between the two approaches (horizontal axis) and the minimal distances
between traps (vertical axis) are modified. (b) shows the 
transfer efficiency as a function
of $t_{\text{Delay}}$ (horizontal axis) and 
of the amplitude of a shaking $a_{\rm Shake}$ in the positions of the outer traps (vertical axis)
with $\omega_{Shake}=10^{-2}\omega_x$. 
For $a_{\text{Shake}}>0$ the shaking of the outer traps is in phase,
for $a_{\text{Shake}}<0$ it is out of phase by $\pi$.
}
\label{fig:tlao:stirap:robust}
\end{figure}

The advantage of such a STIRAP-like process, as compared to a direct transport {\it via}
Rabi-type oscillations, is its robustness with respect to the variation of
certain experimental parameters. As shown in \figref{fig:tlao:stirap:robust} (a),
the scheme works for a large range of delay times $t_{\rm Delay}$
and minimum distances $d_{\rm min}$. A
similar robustness is found for variations of, \eg, the duration $t_r$ of
the approaching/separation process and the time $t_i$ for which the traps are
kept at constant distance,
the only requirements being the adiabaticity of the process and
the order of approaching and separating the traps. In an experimental
realization, certainly a shaking of the centers of the trapping potentials
 provides an important source of decoherence. It can, e.g., be caused
by a mechanical vibration of the microlenses, or by variations in the 
laser phases for optical lattices. Here we anticipate a periodic variation
of the distance of the traps with frequency well below the trapping frequency, 
namely $\omega_{\rm Shake}=10^{-2}\omega_x$.
As \figref{fig:tlao:stirap:robust} (b) shows, the transport
efficiency is not significantly degraded even for shaking amplitudes on the order
of a few  percent of the minimal trap distance if the delay time is appropriately chosen. 

\begin{figure}[tbp]
\begin{center}
\includegraphics[width=\columnwidth]{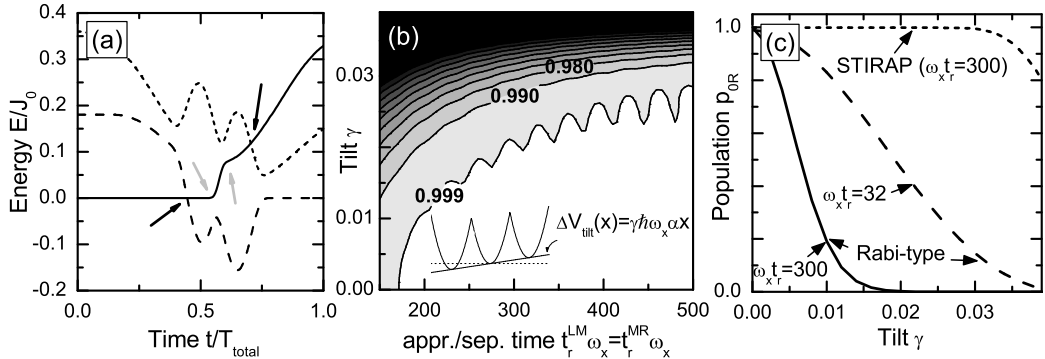}
\end{center}
\caption[]{
Transfer efficiency for an additional tilted potential
$V_{\text{tilt}}(x)=\gamma\;\hbar\omega_x\;\alpha x$,
parametrized by $\gamma$. (a) Temporal variation
of the energy levels obtained from a diagonalization of \eqnref{eqn:ham_lambda}
for the parameters as in \figref{fig:tlao:stirap:single} and $\gamma=0.02$.
Black arrows correspond to diabatic crossings, gray arrows indicate points where
the probability of non-adiabatic transitions is larger. In (b) the transfer
efficiency is plotted as a function of the time $t_r$ needed to approach and separate
the traps and of the tilt; in (c) the dependence of the transfer efficiency on the tilt
of the potentials is compared for STIRAP for the parameters from (b)
with $\omega_x t_r=300$ and for the transfer \via Rabi-type oscillations
between two traps for $\omega_x t_r=300$ ($\omega_x t_i=12$ such that full population transfer occurs for $\gamma=0$)
and $\omega_x t_r=32$ ($\omega_x t_i=25$).
}
\label{fig:tilt}
\end{figure}

A parameter difficult to control in an experiment is the exact horizontal alignment of
the traps. If a slight tilt is present in the potential, then gravity will change the
relative depth of the potential minima.
In this case, as has been reported in Bose-Einstein condensation in a double-trap
potential \cite{tiecke:2003}, after a sufficiently adiabatic evolution,
the atom(s) will eventually be found in the 
trap with lower energy. To allow for a transport to the desired state, the
evolution should be explicitly non-adiabatic \cite{Scharnberg:2004}. Here we will
show that the STIRAP-like transport is within a large range not affected by gravity.
To this aim we add a potential  
\beq
\Delta V_{\rm tilt} (x)=\gamma\hbar\omega_x\alpha x\label{eqn:tlao:tilt}
\eeq
stemming from gravity to the dipolar trapping potential $V(x)$. The 
parameter $\gamma$ determines the slope of the ramp, and we will use $\gamma>0$,
such that the right trap is shifted up in energy with respect to the left one.
For the parameters of our simulations, a value of $\gamma=10^{-2}$
corresponds to a difference in the potential energy of $3\cdot10^{-2}\;\hbar\omega_x$
between the outer traps at the minimal distance. For $\gamma\ll1$ such a tilt
affects only the on-site energies $-\mu_{\alpha\beta}$ in the Hamiltonian (\ref{eqn:ham_lambda}). In
the picture of an optical $\Lambda$-system this corresponds to a shift from the one- as well as from the 
two-photon resonance. In this case there exist no adiabatic path from $\ket{L}$ to 
$\ket{R}$ \cite{fewell:1997}. This is exemplified in \figref{fig:tilt} (a),
which shows the energies of the three eigenstates of the Hamiltonian \eqnref{eqn:ham_lambda}
for the parameters of the STIRAP sequence from
\figref{fig:tlao:stirap:single} and $\gamma=2\times10^{-2}$. To obtain the transport from
the left to the right trap, the process has to be designed as a
combination of diabatic (black arrows) and adiabatic (grey arrows) processes. However,
the conditions to obtain a diabatic crossing at the points indicated by the black arrows 
are usually fulfilled, such that the transfer efficiency is dominated by the
adiabaticity requirement. For this reason, for a given $\gamma$ the
fidelity improves if the approaching and separating of the the traps
is made slower as can be seen from \figref{fig:tlao:stirap:single}. As should be stressed
again, this is in contrast to the Rabi-type transport between two traps, where a faster process
has larger fidelity of the atom ending up in the initially empty trap. \figref{fig:tlao:stirap:robust} (d)
shows a comparison of the two schemes: next to the transfer efficiency
for STIRAP for $\omega_x t_r=300$ as a function of the tile $\gamma$ two curves for
Rabi-type oscillations are shown for a slow ($\omega_x t_r=300$) and for a
fast ($\omega_x t_r=32$) approaching process.

\subsection{CPT-like and EIT-like effects}

From the isomorphism between the Hamiltonians, it is obvious that also
other processes from three-level optics can be exploited
here. The approach sequence can be modified to create spatial superposition states with maximum 
atomic coherence by evolving the mixing angle from $\Theta=0$ to $\Theta=\pi/4$, corresponding to 
a delayed approach of the left trap, but to symmetric separation. This process, reminiscent of 
coherent population trapping (CPT), is similarly robust to the variation of parameters as the STIRAP process,
{\it as long as the symmetry of the separation process is maintained}. A detailed analysis 
can be found in \cite{eckert:2004}, where also an EIT-like process is described.

\subsection{Effects of atom--atom interaction}

To arrive at the Hamiltonian of \eqnref{eqn:ham_lambda}, the system has been
simplified through a restriction to the lowest energy levels and to only a single atom.
Our numerical simulations, which took into account excited states, show that for adiabatic
processes and the potentials considered here the former simplification is justified and allows
to reproduce effects known from $\Lambda$ systems. Excited states however can be exploited
explicitly by starting with the atom in a state different from the ground state \cite{eckert:2004}. On the other hand,
also the restriction to a single particle can be released. For sufficiently low temperatures,
interaction between bosonic atoms is dominated by s-wave scattering, and restricting again to states
$\phi_{\alpha}$, the Hamiltonian describing the system is modified as follows
\cite{calarco:2000,oehberg:1998}:
\beq
\hil=\hil_{Free}+\frac12\sum_{\alpha=L,M,R}
U_{\alpha} \hat{b}_{\alpha}^{\dagger}\hat{b}_{\alpha}^{\dagger}\hat{b}_{\alpha}\hat{b}_{\alpha}.
\eeq
Here $U_{\alpha}=4\pi\hbar \tilde{a}_{\rm sc}\int dx|\phi_{\alpha}(x)|^4$, where $\tilde{a}_{\rm sc}$
is the 1D scattering length which can be changed {\it via} changing the
orthogonal confinement or exploiting a Feshbach resonance. If $|U_{\alpha}|$ is sufficiently
large to separate sectors in energy space with different particle number within one trap, then
interaction allows to move coherently more than one atom at once, or to generate 'Schr{\"o}dinger
cat'-like states through a CPT sequence. On the other hand, starting from a system of three traps
and two bosonic atoms initially in different traps allows to coherently and robustly transport the
'hole', i.e., the empty site. An optical analogue of such a system has been studied in
\cite{mompart:2001}, where coherent
population trapping has been analyzed for two electrons with aligned spins in a three-level system.
\begin{figure}[tbp]
\begin{center}
\includegraphics[width=1.\columnwidth]{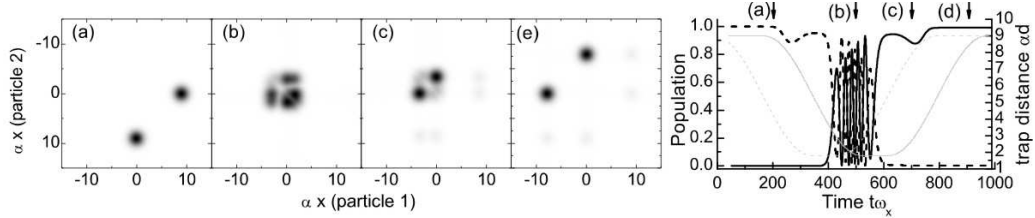}
\end{center}
\caption[]{
Robust and coherent transport of a hole from the left trap
to the right trap in a system of three traps filled with two bosonic
${}^{87}$Rb atoms with scattering length $a_t=106a_0$. Shown are plots of
the two particle probabilities $|\psi(x_1,x_2)|^2$ at four
different times (a-d) as indicated by the arrows in (e).
The initial state is $\ket{\psi(t_{\rm init})}=\ket{0_M}\ket{0_R}+\ket{0_R}\ket{0_M}$,
the parameters are
$t^{LM}_{r}\omega_x=t^{MR}_{r}\omega_x=350$,
$t^{LM}_{i}\omega_x=t^{MR}_{i}\omega_x=100$,
$t_{\rm delay}\omega_x=180$,
$d^{LM}_{max}\alpha=d^{MR}_{max}\alpha=9$, and 
$d^{LM}_{min}\alpha=d^{MR}_{min}\alpha=1.5$.
}
\label{fig:tlao:twoatoms}
\end{figure}
Also in this case a dark state exists which can be interpreted as the dark state of a 'hole'. A similar
effect can be achieved in the atom optical system. As an example, \figref{fig:tlao:twoatoms}
demonstrates the corresponding STIRAP process which moves the hole between the outer traps.

\section{Manipulation of matter waves in guiding structures}

\begin{figure}[tbp]
\begin{center}
\includegraphics[width=1.\columnwidth]{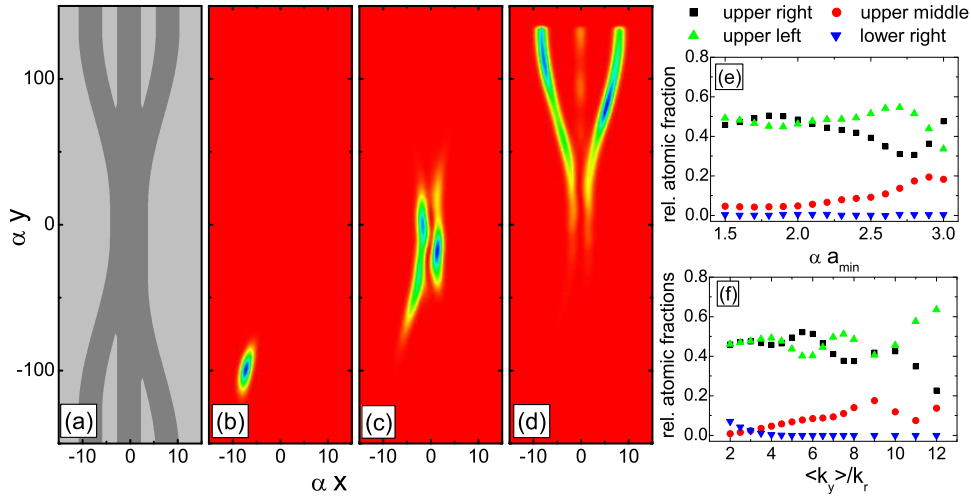}
\end{center}
\caption[]{
(a) Contour plot of the waveguide potential used to split the wave packet
incident in the left arm into a superposition of packets traveling in the left
and the right arm using a CPT--like process. The contour line shown corresponds
to a distance of $3/2$ ground state widths away from the centers of the
waveguides. (b)--(d) show density plots of $|\psi(x,y)|^2$ at times $\omega_x t=20$,
$\omega_x t=60$, $\omega_x t=120$ for a wave packet with mean momentum $\ew{k_y}=3.5k_r$
and initial width $\Delta k_y=k_r$. The minimal distance between waveguides is
$\alpha d_{\rm min}=1.5$. (e) The relative atomic fractions leaving the setup
through the upper and lower exits of the structure as the minimal distance between waveguides
is modified. (f) As (e), but for different mean velocities $\ew{k_y}$.
}
\label{fig:tlao:wgcpt}
\end{figure}

In the previous part we have, in close analogy to the three-level processes
for internal atomic states, manipulated the external wavefunction of a
trapped atom by a temporal variation of the coupling between traps.
We will demonstrate that similar methods allow to manipulate
an atomic wave packet propagating in an appropriately designed 
{\it fixed} guiding structure. We will assume a system of three waveguides oriented
in the direction of the $y$ axes, with $y$-dependent distances (see
\figref{fig:tlao:wgcpt} (a) for an example), and a corresponding Hamiltonian
$H_{\rm free}=(p_x^2+p_y^2)/2m+V(x,y)$. Now, instead of
considering the eigenstates of the 1D potential for each
fixed time $t$, we can compute eigenstates at each position $y$,
and as before combine the states with lowest energy to states
$\phi_{\alpha}(x,y)$, $\alpha\in\{L,M,R\}$ localized around the center of each waveguide.
The full wavefunction can then be decomposed as
$\psi(x,y,t)=\sum_{\alpha}c_{\alpha}(y,t)\phi_{\alpha}(x,y)$, and inserting this expression
into the Schr{\"o}dinger equation gives the following equation for the evolution
of the coefficients $c_{\alpha}(y,t)$ \cite{stenholm:2003}:
\bea
i\hbar\frac{\partial c_{\alpha}}{\partial t}=-\frac{\hbar^2}{2m}\frac{\partial^2c_{\alpha}}{\partial y^2}
+\sum_{\beta=L,M,R}(H_{\alpha\beta}+\frac{\hbar^2}{m}P_{\alpha\beta})c_{\beta}+\frac{\hbar^2}{2m}\sum_{\beta=L,M,R}K_{\alpha\beta}\frac{\partial c_{\beta}}{\partial y}.\label{eq:wgeq}
\eea
Here $H_{\alpha\beta}(y)=\int dx\,\phi^*_{\alpha}(x,y)(p_x^2/2m+V(x,y))\phi_{\beta}(x,y)$ are 
the Hamiltonian matrix elements for fixed position $y$ and
$K_{\alpha\beta}(y)=-\int dx\,\phi^*_{\alpha}(x,y)\partial_y\phi_{\beta}(x,y)$ and
$P_{\alpha\beta}(y)=-\int dx\,\phi^*_{\alpha}(x,y)\partial_y^2\phi_{\beta}(x,y)$ are the 
kinetic and potential couplings, respectively.

For time-dependent trapping potentials, the STIRAP  or CPT sequence was induced
by the counterintuitive temporal ordering of the approaching and separation
processes of the traps. Similarly, in the case of waveguides we will apply such
sequences in space. Thus, to obtain a STIRAP--like transport with the atom initially located in the left
arm, first the right guide is approached to the middle one, then, with an appropriate delay,
the tunneling is also switched on between the middle and the left guide. Finally, tunneling is turned
off in the same order: first between the right and the middle and then between the middle and the left guide.
For a CPT-like process
to split an atomic wave packet coherently between two waveguides, the approaching sequence has
the same counterintuitive order, but the separation is symmetric, see \figref{fig:tlao:wgcpt} (a). Now
the additional coupling terms in \eqnref{eq:wgeq} make the evolution more
complex. Expanding $c_{\alpha}$ into plane waves with momentum $\hbar k$
leads to a diagonal $k^2$-proportional term which accounts for broadening
of the wave packet and to a term proportional to $k\;K_{\alpha\beta}$ which
induces velocity-dependent couplings between the waveguides.
Furthermore, a velocity-independent modification of the couplings is introduced through
the potential couplings $P_{\alpha\beta}$.

To take again into account further effects beyond this illustrative approximation in order to evaluate the 
performance of such processes we have numerically integrated the full 2D
Schr{\"o}dinger equation. We assume to initially have an atomic wave packet located in the left arm,
with a gaussian profile in the direction of the waveguide corresponding to mean momentum
$\ew{k_y}$ and momentum spread $\Delta k_y=k_r$ ($k_r=\sqrt{2m\omega_r/\hbar}$; for the
simulations $\omega_r=\omega_x/6$). In the transverse direction the wave packet
corresponds to the ground state of the potential. \figref{fig:tlao:wgcpt} (b-d) shows 
an example of the time evolution in a structure which generates a splitting of
the wave packet through a CPT--like configuration. During the evolution, the wave packet
strongly broadens in the direction of propagation, but is nevertheless nearly equally
split between the left and right outgoing arms with a negligible amount of
reflection. This splitting is again relatively robust with respect
to the parameters describing the potential, provided the symmetry of the splitting is maintained.
\figref{fig:tlao:wgcpt} (d) shows the change of the atomic fractions in the exit ports of the setup
as the minimal distance of the waveguides is varied. The process is not as perfect as its
counterpart in traps due to the additional couplings present here. Especially the 
velocity-dependent couplings modifying the desired CPT--like process play an important
role: the larger the mean velocity, the stronger the deviation from the equal splitting,
as can be seen from \figref{fig:tlao:wgcpt} (e).

\section{Conclusions}

We have studied the manipulation of the external wavefunction of an atom
in a dipole potential consisting of three traps whose coupling
can be changed as a function of time. As we have demonstrated, such a system,
restricted to the lowest eigenstates, constitutes an analogon to the
extensively studied system of three internal atomic states coupled via 
two external laser fields. This allows to apply concepts as STIRAP, CPT,
or EIT to coherently and robustly manipulate the external atomic wavefunction.
Such processes are of potential interest, e.g., to move around atomic quantum
bits or to create superposition states for interferometry.
In particular, we have analyzed the robustness of a STIRAP--like process
allowing transport of the atom between trapping potentials, and we have also
shown that coherent processes are possible for several interacting atoms.

As a different setup, we have studied atomic wave packets propagating in
waveguide potentials, where the time dependence of the trap distances is replaced
by a spatial variation of the distance between waveguides. Due to additional,
partially velocity-dependent couplings, the evolution is more involved and the
transport or splitting processes are not as clean as in the case of traps. Still, a
stronger robustness as for schemes relying only on Rabi-type tunneling
between traps can be achieved, as we have exemplified through demonstrating the coherent
splitting of a wave packet between two arms, a scheme interesting for,
e.g, interferometry.

\section{Acknowledgments}

We dedicate this paper to Bruce Shore on the occasion of his 70th birthday.
This work has been supported by the European Commission through IST project ACQP,
by the DFG (Schwerpunktprogramm 'Quanteninformationsverarbeitung' and SFB 407), by the 
'Innovationsbudget Hessen', by NIST, ARDA, and NSA, and by the MCyT, MEC (Spanish Government) 
and the DGR (Catalan Government) under contracts
BFM2002-04369-C04-02, FIS2005-04627, and 2001SGR00187, respectively. We also acknowledge
support of ESF programme QUDEDIS and EU IP Programme Scala. KE acknowledges further support from
HPC-Europa. We thank W.~Ertmer, H.~Kreutzmann, A.~Sanpera,
and F.~Scharnberg for discussions.


\end{document}